# EEG Foundation Models for BCI Learn Diverse Features of Electrophysiology

Mattson Ogg, Rahul Hingorani, Diego Luna, Griffin W. Milsap, William G. Coon, Clara A. Scholl
*Research and Exploratory Development Department, Johns Hopkins University Applied Physics Laboratory*
*Laurel, Maryland, USA*

*Abstract*—Brain computer interface (BCI) research, as well as increasing portions of the field of neuroscience, have found success deploying large-scale artificial intelligence (AI) pre-training methods in conjunction with vast public repositories of data. This approach of pre-training foundation models using label-free, self-supervised objectives offers the potential to learn robust representations of neurophysiology, potentially addressing longstanding challenges in neural decoding. However, to date, much of this work has focused explicitly on standard BCI benchmarks and tasks, which likely overlooks the multitude of features these powerful methods might learn about brain function as well as other electrophysiological information. We introduce a new method for self-supervised BCI foundation model pre-training for EEG inspired by a transformer-based approach adapted from the HuBERT framework originally developed for speech processing. Our pipeline is specifically focused on low-profile, real-time usage, involving minimally pre-processed data and just eight EEG channels on the scalp. We show that our foundation model learned a representation of EEG that supports standard BCI tasks (P300, motor imagery), but also that this model learns features of neural data related to individual variability, and other salient electrophysiological components (e.g., alpha rhythms). In addition to describing and evaluating a novel approach to pre-training BCI models and neural decoding, this work opens the aperture for what kind of tasks and use-cases might exist for neural data in concert with powerful AI methods.

## I. Introduction

Non-invasive brain computer interfaces (BCIs) supporting restorative therapeutic capabilities [1], [2], [3], [4] and neural decoding have become a central tool for neuroscientific research [5]. Electroencephalography (EEG) has been a popular modality for non-invasive BCI and neural decoding due to its high temporal resolution and accessibility for the general public (good safety profile, minimal contraindications, multiple consumer and research grade systems available). However, EEG is also characterized by poor spatial resolution, a large degree of inter-individual variability and low SNR for the underlying neural activity [6], all of which limit BCI and neural decoding performance [7], [8].

These challenges are compounded by the fact that data for training neural decoders is almost always severely limited [9], [10]. Generating data for training a machine-learning classifier for BCI is typically expensive, time consuming and laborious, requiring large numbers of cued trials that a participant must repeat tens to hundreds of times. This paradigm also almost always means that in-laboratory data collection is required which limits applicability to fairly stereotyped behavioral classes and challenges many real-world applications (see [11] for discussion). While this might impede the use of typical deep-learning methods (which generally require orders of magnitude more data to train), some model architectures have been cleverly designed to operate in this data-limited regime [12], [13]. While these models have improved BCI performance, their supervised training procedures mean they remain yoked to specific stereotyped tasks and thus challenges remain with respect to generalization across datasets, participants, and more naturalistic behaviors [14].

At the same time, data resources are becoming more abundant [15], [16], [17], raising the possibility that fundamental challenges arising from data limitations in BCI could be addressed by pre-training models on larger datasets and then using these models to jump start training in more typical data-limited BCI and neural decoding scenarios. This process of pre-training and transfer-learning has proven immensely effective in other areas, especially when self-supervised methods are used for model pre-training [18], [19], obviating the need to homogenize class labels across tasks, and allowing pre-training to scale over vast datasets. Initial results applying these to BCI [20], [21], [22], [23] and other neural data applications [24] have been promising (see [25] and [10] for reviews). The assumption is that a better representation of electrophysiology could be learned by the model from this larger-scale view of the data (both for target neural activity and also for other physiological signals, confounds and noise sources to be avoided), which will result in features that will be more discriminative when the model is fine-tuned to perform a downstream task given a more limited set of specific labels. However, much of the work to date has been explicitly focused on BCI tasks and benchmarks. Less attention has been paid to the global features that self-supervised foundation models might learn from vast quantities of electrophysiology data. Results from other domains like speech processing [26], [27], [28] and some initial work examining EEG recorded during sleep [24] suggest these models learn diverse features from their pre-training data that could be useful on a wide variety of downstream tasks.

This work was supported by the Joint Science and Technology Office of the Chemical and Biological Defense Program at the Defense Threat Reduction Agency. DISTRIBUTION STATEMENT A. Approved for Public Release: distribution is unlimited. Cleared for Public Release.
Correspondence: mattson.ogg@jhuapl.edu

In this report we adapted a method previously developed for speech processing ("HuBERT" [26]) to self-supervised pre-training of multi-channel EEG decoding models for BCI tasks. We developed a novel formulation of the BCI foundation modeling task that used a masked prediction loss for pseudo-labels discovered from the data, optimized preprocessing for real-time operation, and minimal channel montage for generalizability to different EEG platforms. We carried out a high-level exploration of what information the model learned from the data, finding these models learn features about human electrophysiology that are useful for standard BCI tasks and beyond.

## II. METHODS

### A. Pre-Training Data and Processing

Data from 14,979 participants from the Temple University Hospital EEG corpus [16] were used for model pre-training. EEG recording sessions greater than one minute in duration were used which yielded over 1,104 days of total EEG data. While this is hospital data and not explicitly BCI data, it constitutes a large EEG corpus sufficient to support training large transformer models, and it is commonly used for pre-training BCI foundation models [20], [21], [22].

For each session we retained eight EEG channels (Fz, Cz, C3, C4, P7, P8, Pz, Oz, with O1 substituted for Oz in cases where it was not present), resampled the data to 125 Hz, band pass filtered between 0.1 and 50 Hz, and applied a common average reference. To allow for future use of the model in real-time systems, no normalization or scaling (e.g., z-score normalization of data across sessions) was applied. All data was converted to units of microvolts in order to keep input data values approximately within ±10e1 from unity, which has been shown to improve training (e.g., avoid vanishing gradients) on unnormalized time series. Each session was then segmented into 1-minute long sequences of continuous EEG that were used as inputs for pre-training.

### B. Self-Supervised Pre-Training

We adapted the HuBERT [26] method and our prior work developing foundation models for sleep-EEG [24] for generating labels for self-supervised pre-training. This involves a two-stage process: In Stage 1, pre-training focuses on bootstrapping a model by predicting k-means cluster labels assigned to each data point based on a *spectrogram*; in Stage 2, the model learns to predict k-means cluster labels *directly from the Stage 1 model's embeddings* (representations of the data near the output layer). Conceptually, the first Stage offers the model an onramp into the data domain by providing a crude map of its structure to learn from externally-derived landmarks (k-means labels from a spectrogram). The second stage allows the model to refine the map it has begun to learn, but directly from the data, unconstrained by the imposition of any external structure.

In the first stage of pre-training (Stage 1), we generated k-means cluster labels using multi-taper spectrograms calculated for each EEG channel across 1-minute segments. Spectrograms were computed using 4-second windows with a 1-second stride, after which spectrograms from all channels were concatenated. A k-means clustering model was fit using frequency bins across channels at each time step, employing a randomly selected 10% subset of the data. This clustering process resulted in 100 cluster labels, with each 4-second EEG segment assigned one label (57 time-step labels per 1-minute EEG segment, across 100 classes). Subsequently, we trained a transformer model for 20 epochs (batch size: 128), using a learning rate schedule that linearly ramped from 2e-6 to 1e-4 in the initial third of training, then gradually decreased to 1e-9. The pre-training task involved predicting k-means labels for masked portions within each 1-minute EEG segment using a cross-entropy loss. Each time step had an 8% probability of initiating a masked region that spanned 10 consecutive labels. Only these masked segments contributed to the training loss.

In preparation for the second stage of pre-training (Stage 2), embeddings were extracted from the output of the Stage-1 model's transformer layers by projecting data into the model's embedding space. These embeddings were used to fit a new k-means model, again utilizing a random 10% data subset, resulting in a new set of 500 cluster labels. In Stage 2, the transformer model was trained to predict these new cluster labels, substantially increasing the temporal resolution of the prediction task (618 time-step labels per 1-minute EEG segment, across 500 classes). Hyperparameters in Stage 2 otherwise matched those of Stage 1.

The transformer model used here was a larger version of the one used in previous work [24], and was similar in size and configuration to the original "HuBERT-base" model in [26], albeit with 8 channels of input to accommodate the EEG montage, rather than 1-channel of speech audio. Most updates to the original HuBERT architecture were to limit downsampling by the model's initial layers given both the relatively low sample rate of EEG data (125 Hz) compared to audio (16 kHz), and the short time windows of data that would eventually be used for downstream tasks (epochs that are a few seconds long). Specifically, the model first comprised a stack of six one-dimensional convolutional layers (each followed by layer-norm and gelu activation; kernel sizes: 21, 3, 3, 3, 2, 2; strides: 3, 2, 2, 1, 1, 1, all with 512 filters). The output of these layers underwent a linear projection to an embedding size of 768 followed by positional encoding. These embeddings were the input to the following stack of 12 transformer encoder layers (each with 12 transformer heads, a feed-forward dimensionality of 3072, gelu activation and 5% dropout), and is where the mask was applied during pre-training. The transformer output was average pooled to align with the sequence of labels if necessary (i.e., for the first stage of pre-training) and then projected to an embedding of size 256 followed by gelu activation and an output layer (of 100 or 500 for pre-training stages one and two, respectively). In total, the model comprised 96.4 million trainable parameters. See [24] for additional details and a visual depiction of this architecture and approach.

### C. Fine-Tuning Experiments

Quantifying the performance of a foundation model involves fine-tuning it to support specific tasks using labeled segments (i.e., temporal windows) of data. We used three standard benchmark datasets for these analyses. We performed leave-one-participant-out cross-validation experiments for each dataset and task (although leave-one-*run*-out cross-validation was used for participant recognition experiments). All data in the fine-tuning experiments underwent the same pre-processing procedure as the pre-training data except as noted below.

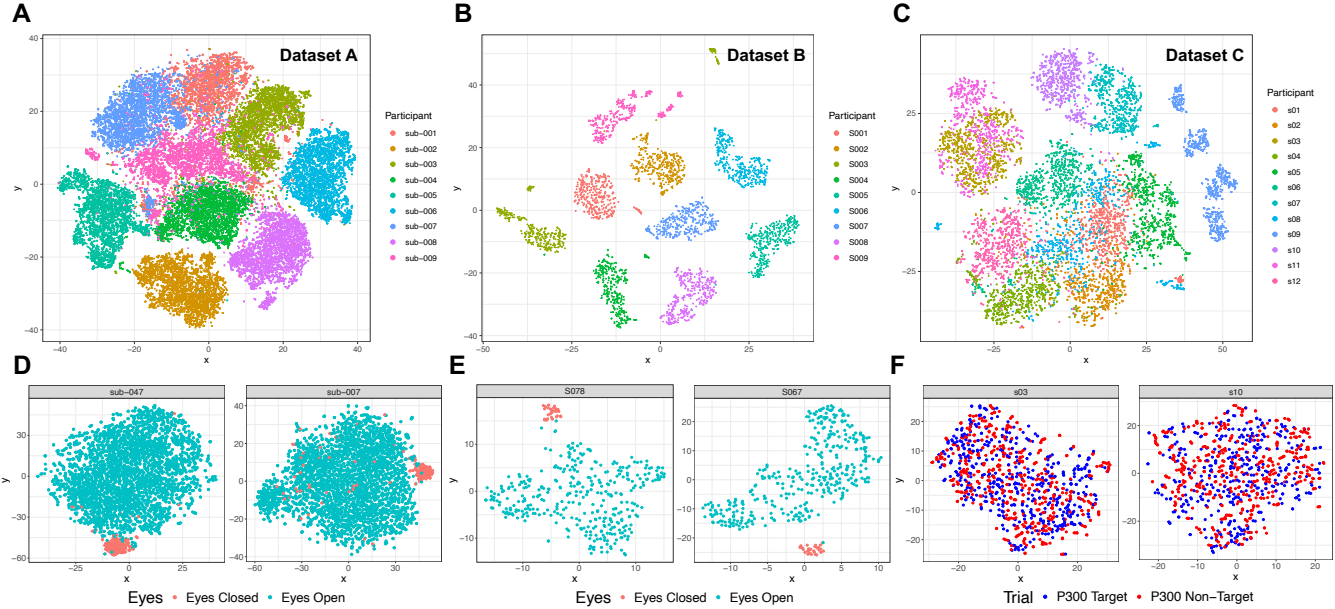

*Fig. 1.* T-SNE visualizations for a subset of participants from each fine-tuning dataset, not seen during pre-training. Activations from the final layer of the transformer were averaged for each trial window prior to any fine-tuning. A through C: T-SNE solutions generated at the dataset-level for subsets of participants in each. D through F T-SNE solutions generated at the individual-level for representative individual participants. Points are all the individual windows of trial data from each participant colored by their participant identifier (A through C) or task (D through F). These visualizations demonstrate the model's sensitivity to individual differences and neural signatures such as alpha rhythms, particularly in eyes-closed rest conditions.

*Dataset A [29]:* EEG data from 55 participants. Dataset consisting of P300 BCI speller trials and RSVP trials. This dataset also included eyes-open and eyes-closed resting state runs. Thirteen (13) total runs per participant. Data were windowed between -200 and 1500-ms around each trial (or every 2-seconds for resting state runs). Non-target trials within ±500-ms of the onset of a target trial were ignored.

*Dataset B [30], [31]:* EEG data from 109 participants. Motor imagery and motor movement tasks for left vs right hand activity ("Task 1" for movement, "Task 2" for imagery) or both-hands vs both-feet activity ("Task 3" for movement, "Task 4" for imagery). This dataset also included eyes-open and eyes-closed resting state runs. Fourteen (14) total runs per participant. Data were windowed between -200 and 4000-ms around each trial (again, every 2-seconds for resting state runs).

*Dataset C [32], [31]:* EEG data from 12 participants. P300 speller trials with no additional resting state runs. Twenty-one (21) total runs per participant. Data were windowed from -200 to 1500-ms around each trial. Non-target trials within ±500-ms of the onset of a target trial were ignored.

For all experiments, models were fine-tuned for 15 epochs and the model corresponding to the epoch with the best performance on a randomly selected validation partition was used for final evaluation of that test partition (batch size: 64 to 256; peak learning rate was 0.0005 with the same linear ramp up and down otherwise). For these experiments, transformer outputs were averaged to one embedding corresponding to each epoch. For each cross-validation fold the model was initialized using the weights learned from the second stage of pre-training. During fine-tuning we either updated all model weights, or just the output layers after the transformer layers (i.e., updated final embedding and output layers, freezing all transformer and convolutional layers). We ran a final set of fine-tuning experiments to assess the value of pre-training by initializing the model without any pre-trained weights and trained that model from scratch (de novo) in each cross-validation fold.

### III. RESULTS

#### A. Visualization of Pre-Trained Model Representations

We first provide high-level exploratory analyses to understand some of the electrophysiological features the model learned. This was carried out via a qualitative t-Distributed Stochastic Neighbor Embedding (T-SNE; [33]) visualization to project pre-trained model embeddings of unseen EEG data into a low dimensional space, *prior to any fine-tuning*. This provides a fairly unbiased assessment of how the model has organized the latent space of EEG data it was trained on. For this, we sent all epochs from a subset of participants in the fine-tuning datasets through the pre-trained model, averaged the transformer embeddings over time for each trial (i.e., window of data, see II.C), and visualized these embeddings using T-SNE solutions generated for each dataset or for each participant.

These T-SNE solutions are displayed in Fig. 1. The model clearly represents participant-level variability within its latent space (Fig. 1A, B, C), which is a prominent feature of neural function [34] and an awareness of this dimension of variability is likely helpful for handling unseen participants and supporting generalization in neural decoding. We also observed clusters corresponding to resting state trials, particularly those where the participant's eyes are closed (Fig. 1D, E). This suggests that the model likely learned to recognize activity related to alpha-rhythms which are a prominent neural response that emerges relatively soon after a participant closes their eyes [35], [36]. Unfortunately, less clear separation is seen among BCI-related trials (e.g., for P300; Fig. 1F).

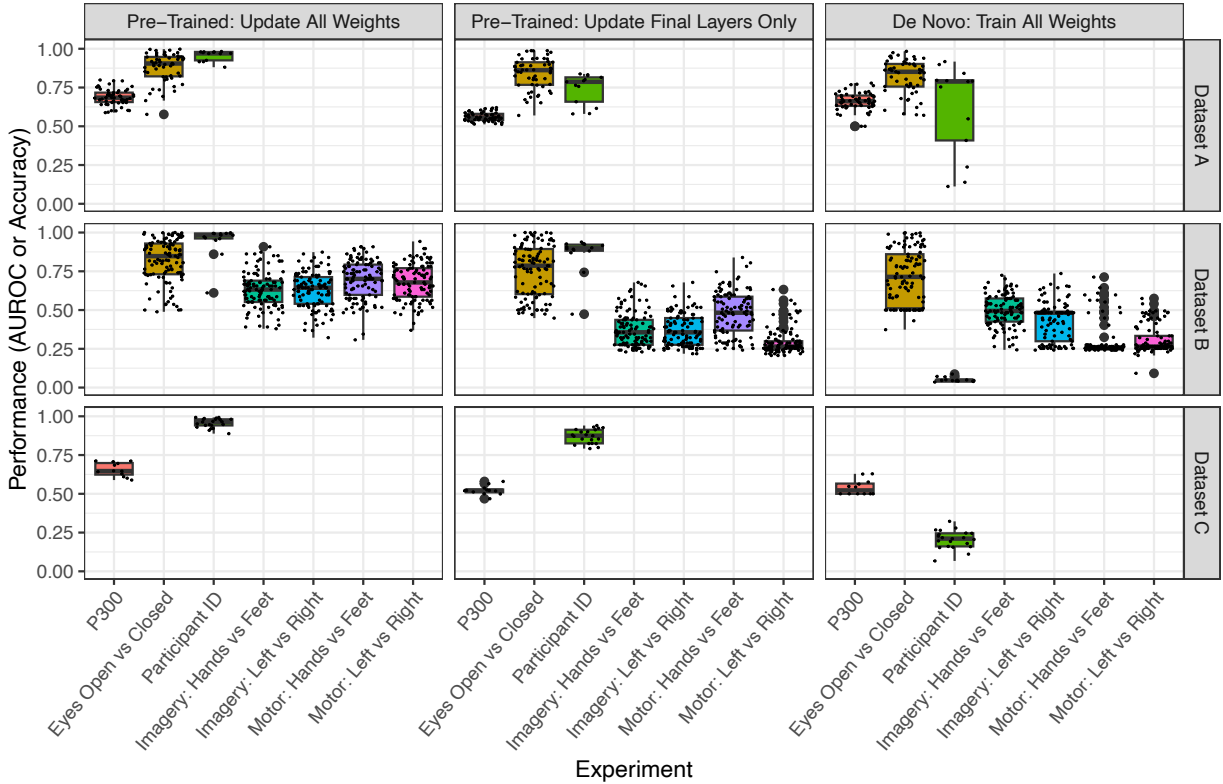

*Fig. 2.* Model performance for each dataset and task. Individual points are the model's performance on each test fold (i.e., for each participant in most cases, or for each run in the case of participant ID). For two-class tasks (where chance = 0.5), AUROC is plotted, in all other cases performance is reported in terms of classification accuracy. For each dataset performance can be visually compared across models by comparing performance within rows. Across datasets, updating all the model's weights yielded the best performance, outperforming both a de novo baseline trained from scratch and a fine-tuned model where only the output layers were updated.

### B. Fine-Tuning Experiments

We quantified the performance of our foundation model using standard BCI benchmark datasets and a leave-one-participant-out cross-validation approach. We also evaluated the model's ability to recognize individuals within each dataset across runs (using leave-one-run-out cross validation), and the accuracy with which the model could detect if the participant's eyes were closed for a resting state run. These results are shown in Table 1 and Fig. 2.

In all cases pre-training the model improved performance on downstream tasks over a (de novo) version of the model trained on each downstream dataset (assessed via non-parametric Wilcoxon rank-sum tests, all $p < 0.001$). Updating the entirety of the model weights during fine-tuning also improved performance over freezing most model layers and updating only portions of the model weights (specifically, the final embedding and output layers; for all comparisons,

Wilcoxon rank-sum tests $p < 0.01$). In general, the model's best performance on BCI tasks is substantially better than chance (all Wilcoxon rank-sum tests, $p < 0.001$) but does not exceed the current state of the art for other models reported in the literature (e.g., [20], though prior work used more EEG channels than the 8 used here and we note that cross validation schemes and other event time-windowing choices or hyperparameters might differ). On non-BCI tasks, both the fully-updated model and frozen model with only output layers updated were able to perform the eyes open or closed and participant recognition tasks extremely well, indicating the model did learn important features of these electrophysiological data, but that more work is required to improve performance for BCI tasks specifically.

TABLE I. Fine-tuning performance

| Dataset | Task | Chance | Update Full Model | Update Output Layers | De Novo Baseline |
|---|---|---|---|---|---|
| Dataset A | P300 | 0.5 | 0.69 | 0.56 | 0.66 |
| Dataset A | Participant ID | 0.08 | 0.96 | 0.74 | 0.62 |
| Dataset A | Eyes Open vs Closed | 0.5 | 0.88 | 0.84 | 0.82 |
| Dataset B | Imagery: Hands vs Feet | 0.33 | 0.63 | 0.37 | 0.5 |
| Dataset B | Imagery: Left vs Right | 0.33 | 0.63 | 0.37 | 0.43 |
| Dataset B | Motor: Hands vs Feet | 0.33 | 0.7 | 0.48 | 0.29 |
| Dataset B | Motor: Left vs Right | 0.33 | 0.68 | 0.29 | 0.31 |
| Dataset B | Participant ID | 0.01 | 0.95 | 0.86 | 0.05 |
| Dataset B | Eyes Open vs Closed | 0.5 | 0.82 | 0.75 | 0.72 |
| Dataset C | P300 | 0.5 | 0.66 | 0.52 | 0.54 |
| Dataset C | Participant ID | 0.02 | 0.96 | 0.87 | 0.21 |

Performance across tasks and datasets. For two-class tasks (chance = 0.5) AUROC is used, otherwise accuracy is reported

## IV. Discussion

We report a novel approach for BCI foundation model pre-training using EEG. This method is based on previous work in speech [26] and EEG recorded during sleep [24]. Despite a minimal EEG montage (just 8 channels) and pre-processing pipeline (no normalization) our model achieves above chance performance on many standard BCI benchmark tasks in the difficult regime of testing the model on unseen participants during fine-tuning. We also observe strong (in some cases near ceiling) performance on novel non-BCI tasks, indicating the model learned diverse neurophysiological features from EEG during pre-training, including potentially recognizing alpha-rhythms and patterns of individual variability within EEG.

While this model can be fine-tuned to achieve good performance on BCI tasks, the model does not exceed state of the art results for these benchmarks. One reason for this could be that the TUEG data used to pre-train the model is recorded from participants in a hospital setting and may not capture the kinds of features that are prominent in BCI (see [10] for discussion). Future work will examine pre-training these models with more diverse BCI or task-specific data to improve downstream task performance in these domains.

More work is also needed to understand whether any other, potentially even non-neural, physiological information is learned by this model from the EEG data. For example, it is likely that eye-movement or other artifacts are encoded by the model. Awareness of these artifacts within a foundation model could be useful for future applications improving robustness, pre-processing and may even support novel control applications.

We demonstrate that self-supervised pre-training on a large corpus of EEG data allows the model to learn many diverse features about neurophysiology. These features may be relevant for additional downstream tasks. One use could be for clinical diagnostics or monitoring alertness, which could be effective given the clinical nature of the pre-training data. This model might also be useful for personalizing BCI interactions given that its learned latent space can account for individual variability, as reflected in the Participant ID task. We are excited to explore this new frontier in neural decoding as this work develops.